\documentclass[aps,prd,reprint,a4paper,showpacs,nofootinbib,superscriptaddress]{revtex4-2}

\usepackage{bbm}
\usepackage{amsmath}
\usepackage{graphicx}
\usepackage{amssymb}
\usepackage{subfigure}
\usepackage{amssymb}
\usepackage{hyperref}
\usepackage{epstopdf}

\usepackage[utf8]{inputenc}
\hypersetup{
    colorlinks=true,
    linkcolor=red,
    citecolor=blue,
}
\usepackage{color}
\usepackage[T1]{fontenc}
\usepackage{txfonts}
\usepackage{mathrsfs}
\usepackage{latexsym}
\usepackage{epsfig}
\usepackage{epstopdf}
\usepackage{epstopdf}
\usepackage{graphicx}
\usepackage{amssymb}
\usepackage{amsmath}
\usepackage{dcolumn}
\usepackage{bm}
\usepackage{color}
\usepackage{comment}
\usepackage{xcolor}
\usepackage{dsfont}

\usepackage{graphicx}
\usepackage{dcolumn}
\usepackage{bm}
\usepackage{color}
\usepackage{enumitem}
\usepackage{amsmath}
\usepackage{amssymb}

\newcommand{\beq}{\begin{equation}}
\newcommand{\eeq}{\end{equation}}
\newcommand{\bea}{\begin{eqnarray}}
\newcommand{\eea}{\end{eqnarray}}

\usepackage{bbm}
\usepackage{amsfonts}
\usepackage{mathrsfs}
\usepackage{latexsym}
\usepackage{epsfig}
\usepackage{epstopdf}
\usepackage{epstopdf}
\usepackage{graphicx}
\usepackage{amssymb}
\usepackage{amsmath}
\usepackage{dcolumn}
\usepackage{bm}
\usepackage{color}
\usepackage{comment}
\usepackage{xcolor}

\usepackage{graphicx}
\usepackage{dcolumn}
\usepackage{bm}
\usepackage{color}
\usepackage{enumitem}
\usepackage{amsmath}
\usepackage{amssymb}

\usepackage{bbm}
\usepackage{amsfonts}
\usepackage{mathrsfs}
\usepackage{latexsym}
\usepackage{epsfig}
\usepackage{epstopdf}
\usepackage{epstopdf}
\usepackage{graphicx}
\usepackage{amssymb}
\usepackage{amsmath}
\usepackage{dcolumn}
\usepackage{bm}
\usepackage{color}
\usepackage{comment}
\usepackage{xcolor}

\begin{document}

\title{Light Neutralino Dark Matter in a Supersymmetric Pati--Salam Framework}

\author{Ali Muhammad}
\email{alimuhammad@phys.qau.edu.pk}
\affiliation{CAS Key Laboratory of Theoretical Physics, Institute of Theoretical Physics, Chinese Academy of Sciences, Beijing 100190, China}
\affiliation{School of Physical Sciences, University of Chinese Academy of Sciences, No. 19A Yuquan Road, Beijing 100049, China}

\author{Imtiaz Khan}
\email{ikhanphys1993@gmail.com}
\affiliation{Department of Physics, Zhejiang Normal University, Jinhua, Zhejiang 321004, China}
\affiliation{Research Center of Astrophysics and Cosmology, Khazar University, Baku, AZ1096, 41 Mehseti Street, Azerbaijan}
\affiliation{Zhejiang Institute of Photoelectronics, Jinhua, Zhejiang 321004, China}

\author{Tianjun Li}
\email{tli@itp.ac.cn}

\affiliation{School of Physics, Henan Normal University, Xinxiang 453007, P. R. China}

\author{Shabbar Raza}
\email{shabbar.raza@fuuast.edu.pk}
\affiliation{Department of Physics, Federal Urdu University of Arts, Science and Technology, Karachi 75300, Pakistan}

\author{Mussawir Khan} 
\email{mussawirkhan@ihep.ac.cn}
\affiliation{State Key Laboratory of Particle Astrophysics, Institute of High Energy Physics, Chinese Academy of Sciences, Beijing 100049, China}
\affiliation{University of Chinese Academy of Sciences, Beijing 100049, China}

\begin{abstract}
We investigate the low-energy phenomenology of the Minimal Supersymmetric Standard Model (MSSM) arising from the supersymmetric
$SU(4)_C \times SU(2)_L \times SU(2)_R$ Pati--Salam framework, focusing on neutralino dark matter in the bulk annihilation and Higgs/$Z$-funnel regions. Using a comprehensive parameter-space scan consistent with radiative electroweak symmetry breaking and a neutralino lightest supersymmetric particle (LSP), we analyze the impact of current collider, flavor, cosmological, and dark matter direct-detection constraints for both signs of the Higgsino mass parameter $\mu$. To isolate genuine bulk annihilation from coannihilation-dominated regions, we impose the conservative mass-splitting condition
$\mathcal{R}_{\tilde{\phi}} \equiv (m_{\tilde{\phi}}-m_{\tilde{\chi}_1^0})/m_{\tilde{\chi}_1^0}\gtrsim 10\%$.
We identify a viable bulk region characterized by a bino-like neutralino LSP and a light right-handed stau as the next-to-lightest supersymmetric particle (NLSP). The allowed bulk solutions satisfy all present experimental constraints, including LHC sparticle searches, flavor observables, and the \textit{Planck} 2018 relic-density bound, and predict robust upper bounds
$m_{\tilde{\chi}_1^0}\lesssim 110~{\rm GeV}$ and
$m_{\tilde{\tau}_1}\lesssim 120~{\rm GeV}$. Our this parameter space lies within the projected reach of future lepton colliders such as CEPC and FCC-ee. We further analyze the Higgs- and $Z$-funnel regions and show that current direct-detection limits strongly constrain light Higgsino-assisted resonance solutions for $\mu>0$. In contrast, for $\mu<0$, destructive interference in the Higgs-mediated spin-independent scattering amplitude suppresses the direct-detection cross section, allowing a narrow but viable $Z$-funnel region to survive below the projected sensitivity of the 1000-day LUX-ZEPLIN exposure. These results identify the negative-$\mu$ realization of the supersymmetric Pati--Salam framework as a particularly predictive and experimentally testable scenario, with promising prospects for both upcoming dark matter searches and future high-energy lepton colliders.
\end{abstract}

\maketitle
\section{Introduction}

Supersymmetry (SUSY) continues to stand as one of the most compelling and theoretically well-motivated extensions of the Standard Model (SM). Beyond providing a symmetry between bosonic and fermionic degrees of freedom, softly broken SUSY frameworks offer a natural stabilization of the electroweak scale against large radiative corrections~\cite{Witten:1981nf}. In addition, supersymmetric gauge theories significantly improve the unification of gauge couplings at high energies~\cite{Dimopoulos:1981yj,Amaldi:1991cn,Ellis:1990wk}, thereby strengthening the case for grand unification. A further remarkable feature of the Minimal Supersymmetric Standard Model (MSSM) is the existence of a stable weakly interacting massive particle (WIMP) when $R$-parity is conserved, providing an attractive dark matter candidate consistent with cosmological observations~\cite{Jungman:1995df,Olive:2003iq,Drees:2005bx}. Moreover, radiative corrections in the MSSM predict an upper bound on the lightest CP-even Higgs boson mass,
$
m_h \lesssim 135~{\rm GeV},$
which is in excellent agreement with the experimentally observed Higgs boson at the LHC with mass around $125~{\rm GeV}$~\cite{Slavich:2020zjv,ATLAS:2012yve,CMS:2012qbp}. This remarkable consistency places strong constraints on supersymmetric parameter space and motivates detailed phenomenological investigations of well-motivated ultraviolet completions.

Among such ultraviolet frameworks, supersymmetric grand unified theories based on the Pati--Salam gauge structure,
$SU(4)_C \times SU(2)_L \times SU(2)_R$, provide an elegant intermediate realization between the SM and larger unification groups such as $SO(10)$. In these constructions, quarks and leptons are unified within common multiplets, and the resulting gauge structure naturally leads to non-universal gaugino masses at the grand unification scale. In particular, the hypercharge gaugino mass satisfies the relation
\begin{equation}
M_1=\frac{3}{5}M_2+\frac{2}{5}M_3,
\end{equation}
which plays a crucial role in determining the low-energy neutralino and slepton spectra. Such non-universality significantly enriches the phenomenology, enabling viable dark matter scenarios while simultaneously accommodating approximate third-generation Yukawa unification~\cite{You:2014vea,Ananthanarayan:1992cd,Shafi:1991rs,Hall:1993gn,Ananthanarayan:1994qt,Rattazzi:1995gk,Blazek:1996yv,Chkareuli:1998wi,Baer:2000jj,Baer:2010ny,Ajaib:2013kka,Ajaib:2013uda,Gogoladze:2009ug,Gogoladze:2009bn,Gogoladze:2010fu,Gomez:2020gav,Djouadi:2022gws}. In this context, the sign of the Higgsino mass parameter $\mu$ becomes particularly relevant, as it influences threshold corrections, electroweakino mixing patterns, and dark matter scattering amplitudes.

Despite the absence of direct evidence for supersymmetry at the LHC, large regions of the MSSM parameter space remain phenomenologically viable. Current collider searches have pushed the masses of colored superpartners, such as gluinos and first-generation squarks, into the multi-TeV regime~\cite{ATLAS:2019vcq,ATLAS:2020dsf,ATLAS:2019gdh,CMS:2019twi,CMS:2019xjf}. In contrast, electroweak states, including neutralinos and sleptons, may still reside near the electroweak scale. This possibility is particularly important for bino-like neutralino dark matter, which generically tends to overclose the Universe in the absence of additional annihilation mechanisms~\cite{Drees:1992am, Khan:2025ibo, Khan:2023ryc}. Viable relic abundance can nevertheless be achieved through several well-established mechanisms, including coannihilation with nearly degenerate superpartners, resonant annihilation via Higgs or $Z$ boson exchange, or mixed bino--Higgsino/wino scenarios.

Among these possibilities, the so-called bulk annihilation region is particularly appealing. In this regime, the correct relic abundance is obtained predominantly through $t$- and $u$-channel slepton-mediated annihilation processes, without requiring resonance enhancement or finely tuned mass degeneracies~\cite{King:2006tf}. However, this region is highly constrained in conventional supersymmetric constructions due to stringent bounds from slepton searches and dark matter direct-detection experiments. In particular, the latest results from XENONnT and LUX-ZEPLIN (LZ) impose severe limits on the spin-independent neutralino--nucleon scattering cross section, significantly restricting light neutralino dark matter scenarios~\cite{XENON:2023cxc,LZ:2018qzl,LZ:2022lsv,LZ:2024zvo}. These constraints are especially relevant for Higgs- and $Z$-funnel regions, where enhanced couplings to Higgs and electroweak gauge bosons can lead to detectable scattering rates.

An additional and phenomenologically important aspect arises from the sign of the Higgsino mass parameter $\mu$. Beyond its impact on electroweak symmetry breaking and supersymmetric contributions to $\Delta a_\mu$, the sign of $\mu$ crucially affects the interference structure of neutralino--Higgs interactions governing spin-independent scattering. In particular, for $\mu<0$, destructive interference between CP-even Higgs exchange amplitudes can substantially suppress the effective neutralino--nucleon coupling~\cite{Ellis:2000ds,Baer:2003jb}, thereby opening regions of parameter space that are otherwise excluded for $\mu>0$. This effect is particularly relevant in scenarios with light Higgsino-assisted $Z$- and Higgs-resonance annihilation.

Motivated by these considerations, we perform a comprehensive study of neutralino dark matter within the supersymmetric Pati--Salam framework, focusing on the bulk annihilation region as well as Higgs- and $Z$-funnel solutions for both signs of $\mu$. To ensure a clean separation between genuine bulk annihilation and coannihilation-dominated regimes, we impose the conservative mass-splitting criterion
$
\mathcal{R}_{\tilde{\phi}} \equiv {m_{\tilde{\phi}}-m_{\tilde{\chi}_1^0}}/{m_{\tilde{\chi}_1^0}} \gtrsim 10\%,$
where $\tilde{\phi}=\tilde{e}_R,\tilde{\tau}_1$. Under this requirement, we identify a consistent and phenomenologically viable region characterized by a bino-like lightest neutralino and a light right-handed stau acting as the next-to-lightest supersymmetric particle (NLSP). The resulting spectrum satisfies
$m_{\tilde{\chi}_1^0}\lesssim 110~{\rm GeV}$ and $m_{\tilde{\tau}_1}\lesssim 120~{\rm GeV}$, while configurations with a right-handed selectron NLSP are found to be strongly disfavored by current ATLAS constraints. We further demonstrate that Higgs-funnel scenarios are largely excluded by present direct-detection limits, whereas a narrow but phenomenologically consistent $Z$-funnel region survives only for $\mu<0$, with spin-independent cross sections lying just below the projected sensitivity of the 1000-day LZ exposure. These results highlight the negative $\mu$ realization of the supersymmetric Pati-Salam framework as a particularly predictive and experimentally testable scenario, with promising discovery prospects at future dark matter experiments and high-energy lepton colliders such as CEPC and FCC-ee~\cite{FCC:2018byv,FCC:2018evy,CEPCStudyGroup:2018ghi,Khan:2025ibo,Khan:2023ryc}.

\section{Model Parameters, Scanning Procedure, and Experimental Constraints}
\label{sec:scan}

We investigate the low-energy phenomenology of the supersymmetric
$SU(4)_C \times SU(2)_L \times SU(2)_R$ Pati--Salam framework, assuming the MSSM as the effective theory below the grand unification scale. In this setup, the high-scale theory is fully specified by a set of soft supersymmetry-breaking (SSB) parameters defined at $M_{\rm GUT}$,
\begin{equation}
m_0,\; m_{H_u},\; m_{H_d},\; A_0,\; M_2,\; M_3,\; \tan\beta,\; {\rm sign}(\mu),
\label{params}
\end{equation}
where $m_0$ denotes a universal scalar mass for squarks and sleptons, while $m_{H_u}$ and $m_{H_d}$ correspond to the soft masses of the MSSM Higgs doublets. The parameter $A_0$ represents the universal trilinear coupling, and $M_2$ and $M_3$ are the gaugino masses associated with $SU(2)_L$ and $SU(3)_C$, respectively. 

A characteristic feature of the Pati--Salam gauge structure is the non-universality of the gaugino sector, which fixes the hypercharge gaugino mass through
\begin{equation}
M_1=\frac{3}{5}M_2+\frac{2}{5}M_3.
\end{equation}
All soft parameters are specified at $M_{\rm GUT}$, while $\tan\beta \equiv v_u/v_d$ and the sign of the supersymmetric Higgsino mass parameter $\mu$ are defined at the electroweak scale.

The weak-scale spectrum is computed using the \texttt{ISAJET}~7.85 package~\cite{Baer:1999sp}, which performs a full renormalization group evolution of gauge and Yukawa couplings between the electroweak and grand unification scales. The procedure iteratively solves the coupled RGEs to obtain a consistent supersymmetric mass spectrum and ensures radiative electroweak symmetry breaking (REWSB). Gauge coupling unification is imposed by requiring
$g_1=g_2=g_U$
at $M_{\rm GUT}$, while allowing a controlled $\sim 3\%$ deviation in $g_3$ to account for unknown GUT-scale threshold effects~\cite{Hisano:1992jj,Yamada:1992kv,Chkareuli:1998wi}. Further details of the spectrum calculation and numerical implementation can be found in Refs.~\cite{Paige:2003mg,Baer:2016wkz}.

The fundamental parameter space is explored in the following ranges:
\begin{gather}
0~{\rm TeV} \leq m_{0},\; m_{H_u},\; m_{H_d} \leq 10~{\rm TeV}, \nonumber \\
-5~{\rm TeV} \leq M_2 \leq 0~{\rm TeV}, \nonumber \\
0~{\rm TeV} \leq M_3 \leq 5~{\rm TeV}, \nonumber \\
3 \leq \tan\beta \leq 60, \nonumber \\
-3 \leq A_0/m_0 \leq 3, \nonumber \\
\mu > 0 \ \text{and} \ \mu < 0.
\label{parameterRange}
\end{gather}

To efficiently explore this multi-dimensional space, we employ a Metropolis--Hastings Markov Chain Monte Carlo (MCMC) algorithm~\cite{Allanach:2006fy,Baer:2010ny}, which preferentially samples regions consistent with collider, flavor, and cosmological constraints. Only parameter points satisfying successful REWSB and yielding a neutralino lightest supersymmetric particle (LSP) are retained. This requirement automatically eliminates scenarios with stable charged or colored relics, which are strongly disfavored by cosmology~\cite{ParticleDataGroup:2012pjm}. In addition, we impose current LHC bounds on strongly interacting superpartners, in particular limits on gluino and first- and second-generation squark masses~\cite{ParticleDataGroup:2014cgo}. Low-energy flavor physics further constrains the viable parameter space. We incorporate experimental limits on
${\rm BR}(B_s\rightarrow \mu^+\mu^-)$,
${\rm BR}(b\rightarrow s\gamma)$,
and
${\rm BR}(B_u\rightarrow \tau\nu_\tau)$,
at the corresponding confidence levels~\cite{LHCb:2012skj,HFLAV:2012imy,HFLAV:2010pgm}. These observables provide complementary sensitivity to supersymmetric loop corrections, particularly in regions of large $\tan\beta$. In the dark matter sector, we require the neutralino relic abundance to satisfy the Planck 2018 measurement within the conservative $5\sigma$ range~\cite{Planck:2018nkj}. Together with collider and flavor constraints, the complete set of phenomenological requirements adopted in this analysis is summarized as
\begin{gather}
122~{\rm GeV} \leq m_h \leq 128~{\rm GeV}, \nonumber \\
m_{\tilde{g}} \geq 2.3~{\rm TeV}, \qquad
m_{\tilde{q}} \geq 2.0~{\rm TeV}, \nonumber \\
0.8 \times 10^{-9} \leq
{\rm BR}(B_s \rightarrow \mu^+ \mu^-)
\leq 6.2 \times 10^{-9}
\qquad (2\sigma), \nonumber \\
2.99 \times 10^{-4} \leq
{\rm BR}(b \rightarrow s \gamma)
\leq 3.87 \times 10^{-4}
\qquad (2\sigma), \nonumber \\
0.15 \leq
\frac{{\rm BR}(B_u \rightarrow \tau \nu_\tau)_{\rm MSSM}}
{{\rm BR}(B_u \rightarrow \tau \nu_\tau)_{\rm SM}}
\leq 2.41
\qquad (3\sigma), \nonumber \\
0.114 \leq \Omega_{\rm CDM} h^2 \leq 0.126
\qquad (5\sigma).
\end{gather}

These combined constraints define the phenomenologically viable regions of the parameter space. They serve as the basis for identifying consistent realizations of bino-dominated bulk annihilation as well as Higgs- and $Z$-funnel dark matter solutions within the supersymmetric Pati--Salam framework.

\section{Numerical Results and Discussion}
\label{sec:results}

A well-motivated and generic prediction of low-energy supersymmetry with conserved $R$-parity is the existence of a stable, electrically neutral, and colorless lightest supersymmetric particle (LSP), which naturally provides a weakly interacting massive particle (WIMP) dark matter candidate. In the MSSM framework, the thermal relic abundance of the neutralino LSP can be consistent with cosmological observations across a variety of well-known mechanisms~\cite{Goldberg:1983nd,Ellis:1983ew,Jungman:1995df,Olive:2003iq,Feng:2003zu,Drees:2005bx,Feng:2010gw,Planck:2018nkj}.

\begin{figure}[h!]\centering
\includegraphics[width=7.90cm]{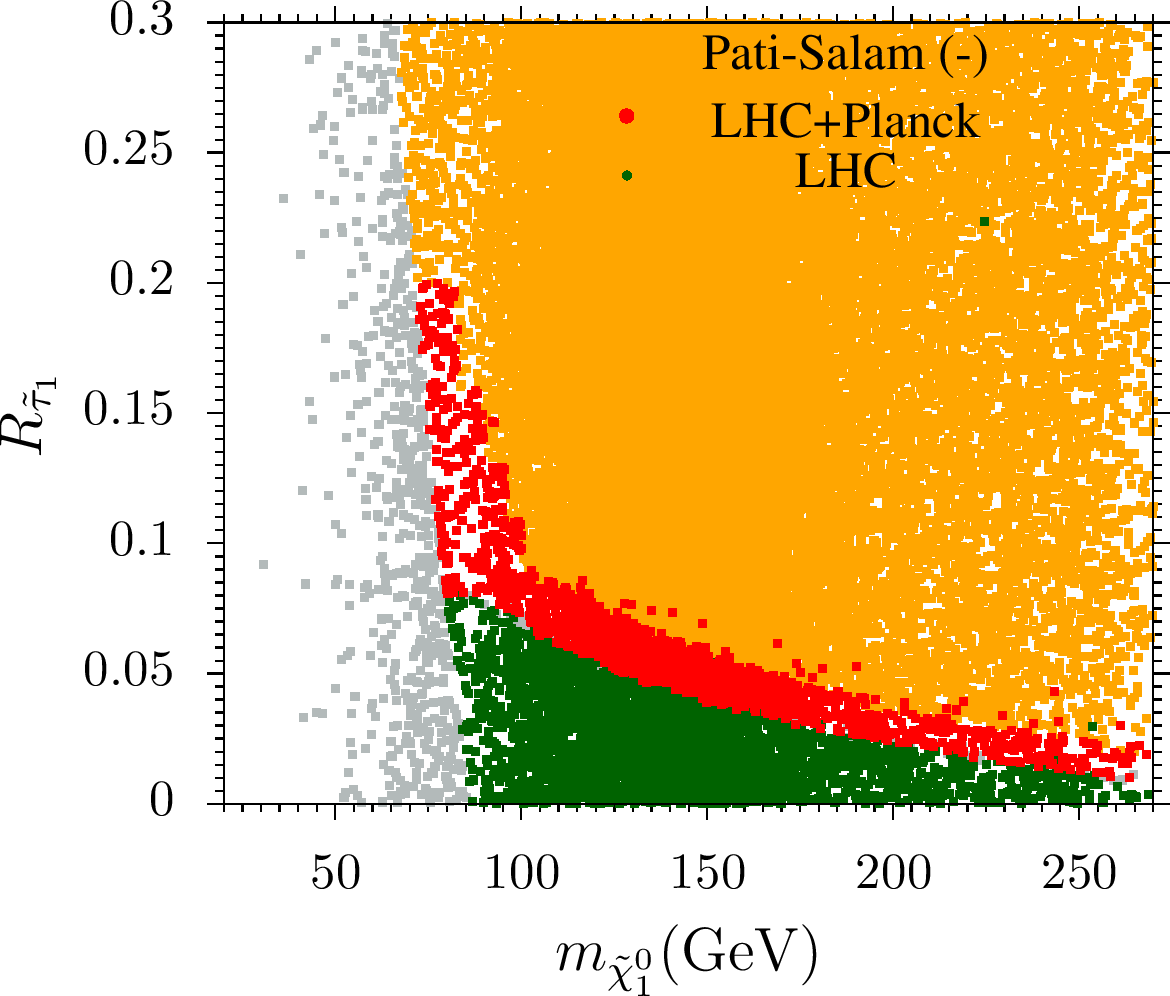}
\caption{\small The grey points correspond to the solutions that satisfy radiative electroweak symmetry breaking (REWSB) and feature a neutralino as the lightest supersymmetric particle (LSP). Among these, the colored subsets—orange, dark green, and red—represent parameter configurations that further satisfy all current experimental constraints, including LEP limits, B-physics observables, Higgs boson mass measurements, and LHC searches for supersymmetric particles. In particular, the orange and dark green points indicate scenarios with over-abundant and under-abundant dark matter relic density, respectively, while the red points correspond to configurations consistent with the observed dark matter abundance. The stau mass splitting is defined as $\mathcal{R}_{\tilde{\tau}_1} \equiv (m_{\tilde{\tau}_1}-m_{\tilde{\chi}_1^0})/m_{\tilde{\chi}_1^0}$, which characterizes the relative mass hierarchy between the lightest stau and the neutralino LSP.}
\label{F1}
\end{figure}

\begin{figure}[h!]\centering 
\includegraphics[width=8.90cm]{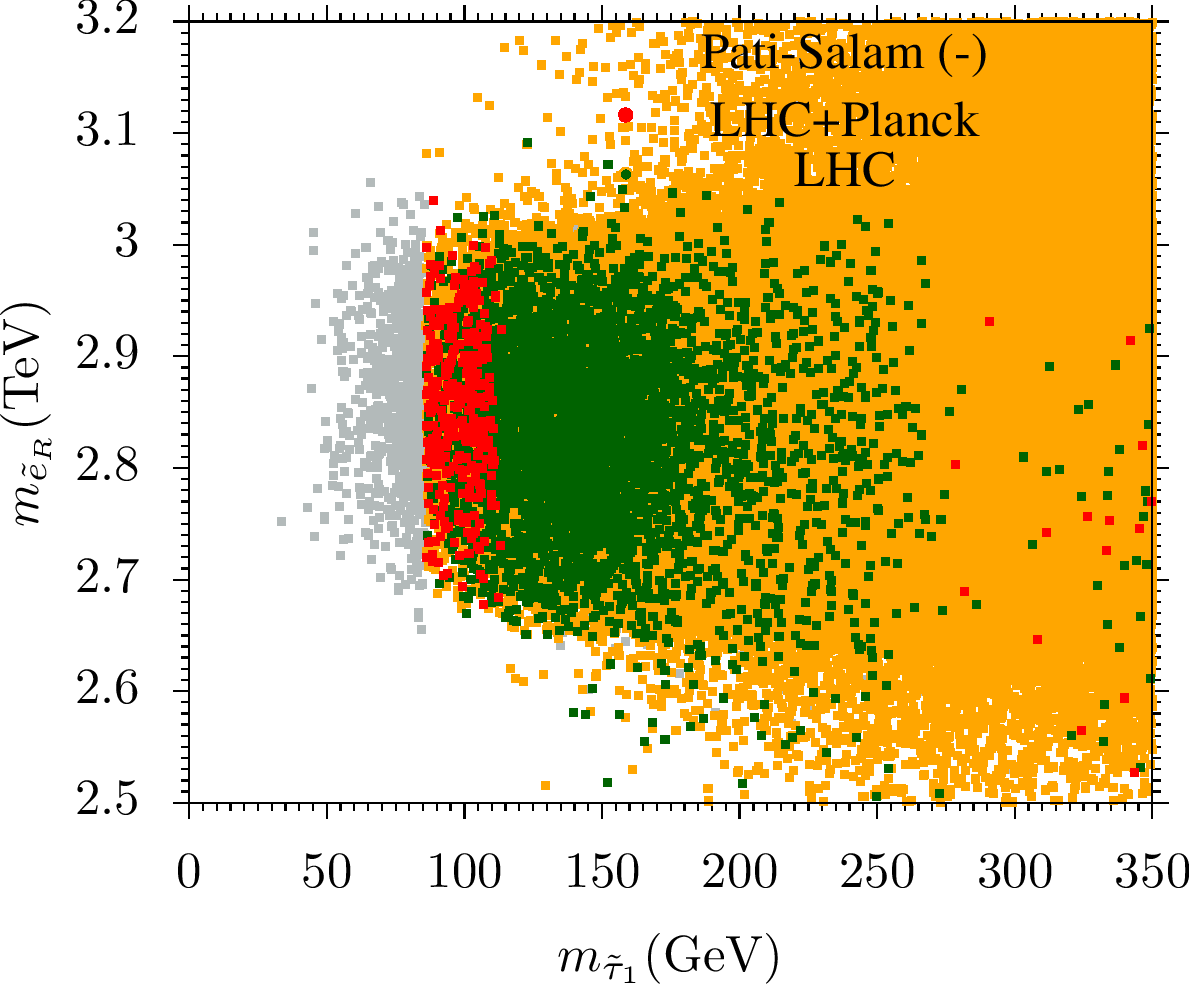}
\caption{\small The gray points represent parameter choices that successfully achieve radiative electroweak symmetry breaking (REWSB) with a neutralino LSP. The colored points denote subsets of the gray points that additionally satisfy the LEP bounds, $B$-physics constraints, Higgs mass requirements, and LHC sparticle search limits. In particular, the orange and dark-green points correspond to overabundant and underabundant dark matter relic densities, respectively, while the red points are consistent with the observed relic abundance and identify the viable bulk region of the parameter space.}
\label{F2}
\end{figure}

\begin{figure}[th!]
\centering \includegraphics[width=8.90cm]{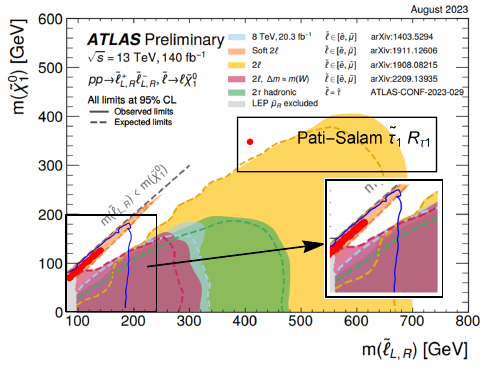}
\caption{\small The allowed bulk parameter space of the supersymmetric Pati--Salam framework is superimposed on the ATLAS August 2023 combined exclusion contours~\cite{ATLAS:2023xco}, which incorporate updated bounds on electroweak slepton pair production~\cite{ATLAS:2014zve,ATLAS:2019lng,ATLAS:2019lff,ATLAS:2022hbt,ATLAS:2023djh}. The surviving points (shown in green) correspond to scenarios with a light stau $\tilde{\tau}_1$ as the next-to-lightest supersymmetric particle (NLSP). In all viable configurations, dark matter is primarily governed by annihilation processes, and the conservative condition on the mass splitting,
$\mathcal{R}_{\tilde{\tau}_1} \equiv (m_{\tilde{\tau}_1}-m_{\tilde{\chi}_1^0})/m_{\tilde{\chi}_1^0} \gtrsim 10\%$, is satisfied, ensuring that stau--neutralino coannihilation effects remain negligible while still reproducing the observed relic density. The ATLAS exclusion (orange region) is mainly derived for selectrons and smuons ($\tilde{e}_{L,R}$, $\tilde{\mu}_{L,R}$), and thus does not strongly constrain the stau sector, since its collider production rates and decay patterns are modified by flavor-dependent mixing and mass ordering. The stau-specific exclusion bounds are indicated separately, and the viable Pati--Salam solutions lie outside the presently excluded regions. The blue line represents the expected reach of the CEPC, showing that a substantial fraction of the phenomenologically interesting parameter space could be probed in future electron--positron collider experiments. The inset provides a magnified view of the low-mass region, with the arrow marking the zoomed-in area of the parameter space.}
\label{F4}
\end{figure}


In general, the lightest neutralino is a linear superposition of bino, wino, and Higgsino gauge eigenstates. While Higgsino- and wino-dominated configurations efficiently annihilate through electroweak interactions into final states such as $W^+W^-$, $ZZ$, $Zh$, $f\bar{f}$, and Higgs boson pairs, the bino-dominated limit is characterized by suppressed annihilation cross sections due to the absence of direct gauge interactions. In this regime, the dominant annihilation channel proceeds via $t$-channel sfermion exchange into fermion--antifermion pairs. This setup defines the so-called \emph{bulk region}, in which the observed relic abundance is obtained without requiring resonance enhancement or coannihilation effects.

To consistently isolate this regime, we focus on highly bino-dominated neutralino configurations and suppress Higgsino- and wino-induced annihilation channels. In addition, we exclude resonance-dominated regions by requiring the neutralino mass to lie sufficiently away from the Higgs and $Z$ poles, i.e. outside the thermally broadened regions around $2m_{\tilde{\chi}_1^0}\simeq m_Z, m_h$, and far from heavy Higgs resonances. 

To eliminate coannihilation effects, we impose a conservative mass-splitting condition,
$
\mathcal{R}_{\tilde{\phi}} \equiv {m_{\tilde{\phi}}-m_{\tilde{\chi}_1^0}}/{m_{\tilde{\chi}_1^0}} \gtrsim 10\%,$
where $\tilde{\phi}$ denotes right-handed sleptons. In particular, we require this condition for both the selectron and the lightest stau states, ensuring that the relic abundance is not affected by coannihilation channels. This selection isolates a regime in which bino annihilation is dominantly controlled by light slepton exchange. We further observe the typical mass hierarchy emerging in the viable parameter space,
$
m_{\tilde{\chi}_1^0}<m_{\tilde{\tau}_1}<m_{\tilde{e}_R}=m_{\tilde{\mu}_R},$
which reflects the interplay of renormalization group running and the underlying Pati--Salam boundary conditions.

Our scan indicates that the requirement $\mathcal{R}_{\tilde{\tau}_1}\gtrsim 10\%$ restricts the neutralino mass to
$m_{\tilde{\chi}_1^0}\lesssim 120~\mathrm{GeV}$.
In addition, scenarios in which the right-handed selectron becomes the next-to-lightest supersymmetric particle are strongly constrained by current LHC searches for soft leptons~\cite{ATLAS:2019lng}. Consequently, the phenomenologically viable realization of the bulk region is characterized by a bino-like LSP accompanied by a comparatively light stau, while first- and second-generation sleptons remain significantly heavier. Figure~\ref{F1} shows the viable parameter space in the $\mathcal{R}_{\tilde{\tau}_1}$--$m_{\tilde{\chi}_1^0}$ plane. The red points correspond to the bulk region satisfying all collider, flavor, and relic-density constraints. The presence of a sizable red region demonstrates that the slepton-mediated bulk annihilation mechanism remains robust within the supersymmetric Pati--Salam framework, even after imposing all current experimental constraints.

The complementary slepton mass plane is shown in Fig.~\ref{F2}. The viable bulk solutions exhibit a characteristic hierarchical structure in the slepton sector, with
$
m_{\tilde{\tau}_1}\lesssim 120~\mathrm{GeV},\, m_{\tilde{e}_R}\lesssim 3~\mathrm{TeV}$.
This hierarchy arises naturally from renormalization group evolution in the presence of non-universal gaugino boundary conditions at the grand unification scale. As a result, the spectrum features a light stau NLSP, while the selectron and smuon states are driven to significantly higher masses, thereby evading current direct LHC constraints while maintaining efficient slepton-mediated annihilation in the early Universe.

\begin{figure*}[th!]
\centering 
\includegraphics[width=8.90cm]{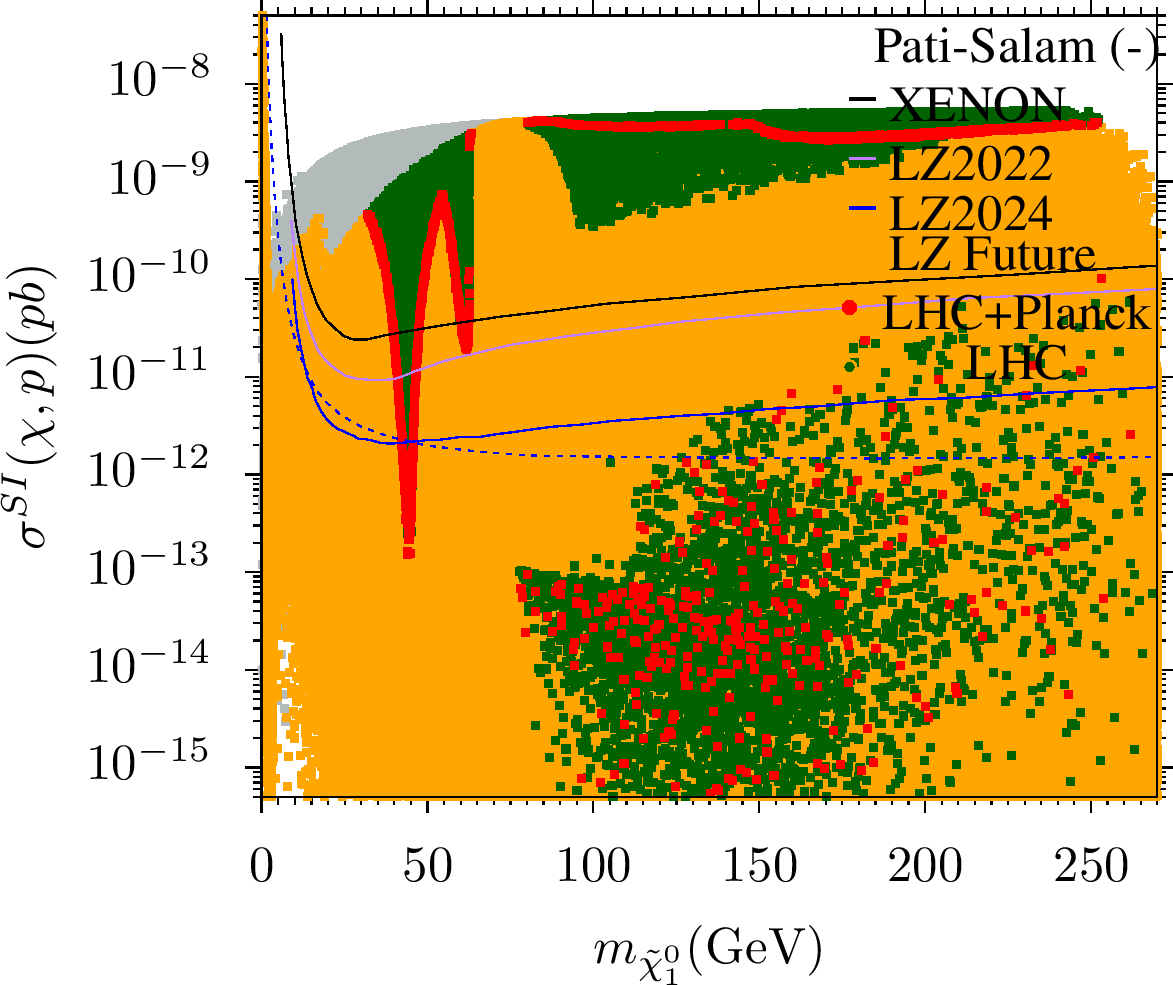}
\centering \includegraphics[width=8.90cm]{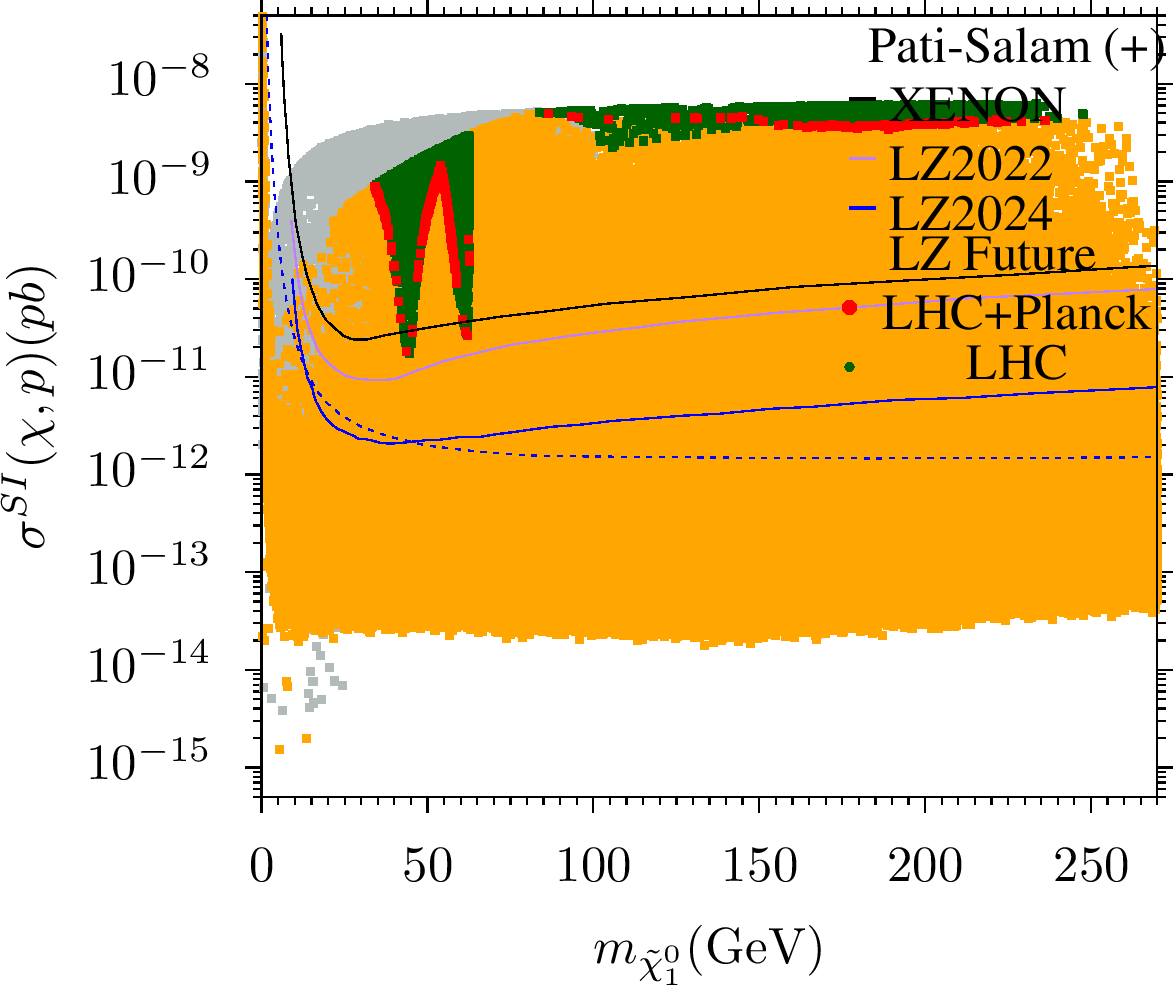}
\centering \includegraphics[width=8.90cm]{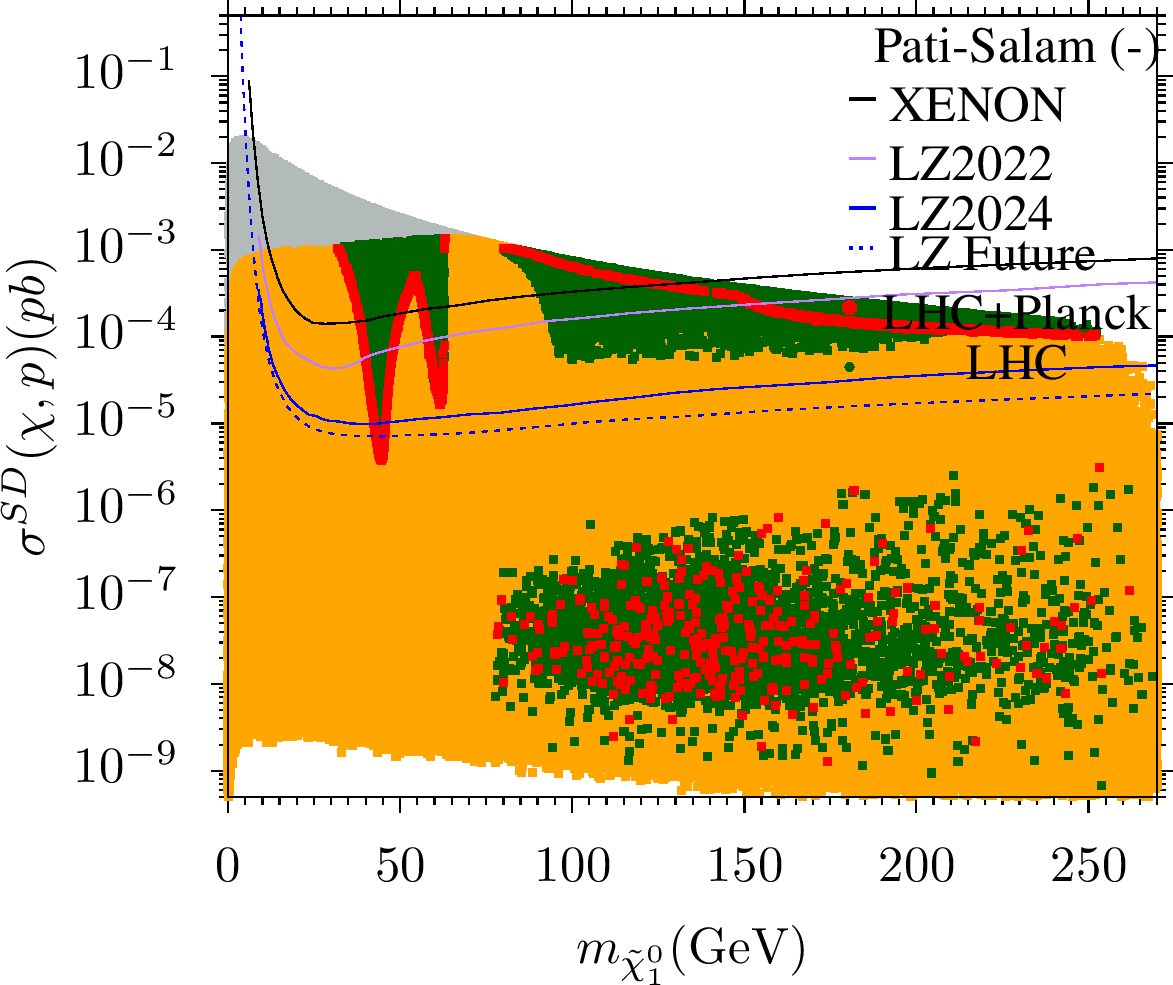}
\centering \includegraphics[width=8.90cm]{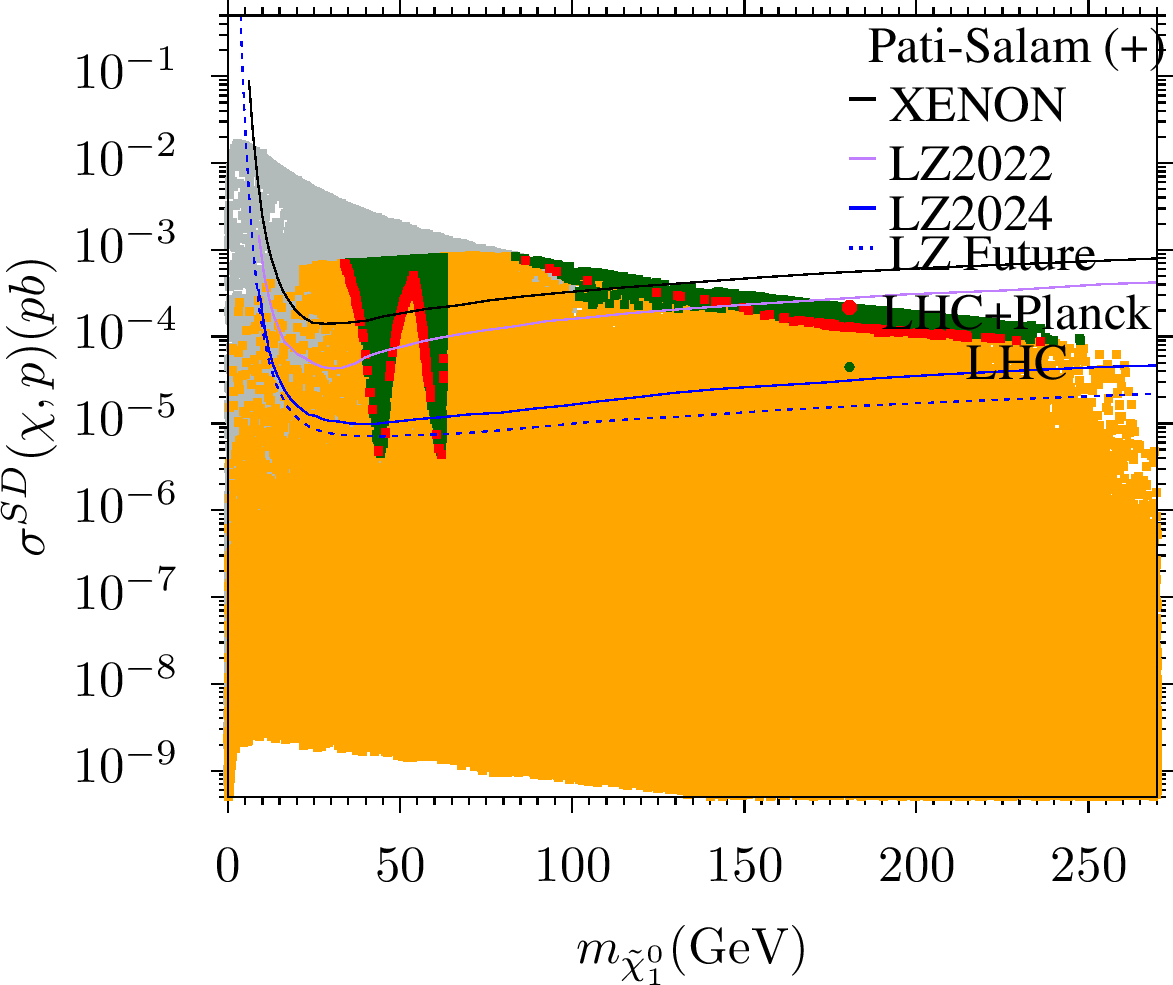}    	\caption{\small The gray points represent parameter choices that successfully achieve radiative electroweak symmetry breaking (REWSB) with a neutralino LSP. The colored points denote subsets of the gray points that additionally satisfy the LEP bounds, $B$-physics constraints, Higgs mass requirements, and LHC sparticle search limits. In particular, the orange and dark-green points correspond to overabundant and underabundant dark matter relic densities, respectively, while the red points are consistent with the observed relic abundance. The upper-left (upper-right) panel shows the spin-independent neutralino--proton scattering cross section, $\sigma_{\rm SI}$, as a function of the neutralino mass for the $\mu<0$ ($\mu>0$) scenario. Similarly, the lower-left (lower-right) panel displays the corresponding spin-dependent scattering cross section, $\sigma_{\rm SD}$, for $\mu<0$ ($\mu>0$). The results are compared with the current exclusion limits from XENONnT (solid black curve)~\cite{XENON:2023cxc} and LZ (solid green and blue curves corresponding to the 2022 and 2024 data sets, respectively). The projected sensitivity of the 1000-day LZ exposure is indicated by the dotted blue curve~\cite{LZ:2018qzl,LZ:2022lsv,LZ:2024zvo}.}
\label{F5}
\end{figure*}

Monojet searches at the LHC~\cite{ATLAS:2020syg,ATLAS:2021kxv} provide stringent constraints on scenarios in which dark matter is produced in association with hard initial-state radiation. However, such analyses are only weakly sensitive to the bino--slepton bulk region considered in this work. The underlying reason is twofold: first, the nearly pure bino nature of the lightest neutralino suppresses direct production rates, and second, electroweak slepton pair production dominates the relevant collider signatures rather than large missing-energy recoil topologies targeted by monojet analyses. Consequently, the efficiencies of monojet searches are significantly reduced in the presence of compressed electroweak spectra and soft final-state leptons. As a result, current monojet searches remain less constraining than direct slepton searches for the parameter space of interest. Figure~\ref{F4} summarizes the current ATLAS slepton and chargino constraints~\cite{ATLAS:2019lng,ATLAS:2023xco,ATLAS:2019lff,ATLAS:2022hbt}, overlaid with the viable bulk solutions of the supersymmetric Pati--Salam model. Although the compressed spectra push much of this parameter space beyond the present LHC sensitivity, future $e^+e^-$ colliders such as FCC-ee~\cite{FCC:2018byv,FCC:2018evy} and CEPC~\cite{CEPCStudyGroup:2018ghi} offer promising discovery prospects. Recent studies of direct slepton pair production at CEPC, based on full Monte Carlo simulations at $\sqrt{s}=360~\mathrm{GeV}$ with an integrated luminosity of $1.0~\mathrm{ab}^{-1}$~\cite{Lyu:2025wjn}, demonstrate significantly enhanced sensitivity to compressed slepton spectra. In particular, CEPC can probe regions where sleptons are nearly degenerate with the LSP, which remain challenging for hadron colliders. The projected CEPC sensitivity, shown by the blue curve in Fig.~\ref{F4}, indicates that a substantial fraction of the Pati--Salam bulk region lies within the reach of future lepton-collider experiments. The projected sensitivities of ATLAS~\cite{ATLAS:2018diz} and CMS~\cite{CMS:2018imu} for stau pair production have also been investigated under the assumption of $\mathrm{BR}(\tilde{\tau}_i\to \tau\tilde{\chi}_1^0)=100\%$. In principle, distinguishing left-handed from right-handed stau production may be possible through differences in the energy distributions of the final-state tau leptons. In this work, we focus on the theoretically motivated hierarchy $m_{\tilde{\tau}_R}\ll m_{\tilde{\tau}_L}$, for which the production cross section of the right-handed stau is generally smaller than that of the left-handed state. Since the simplified models adopted in current collider searches do not always reflect this hierarchy, dedicated analyses of right-stau dominated scenarios remain important for future studies.\footnote{The sensitivity of the High-Luminosity LHC (HL-LHC) to the bulk region considered here remains uncertain. This is particularly relevant in light of the analysis in Ref.~\cite{CidVidal:2018eel}, which assumes $m_{\tilde{\tau}_R}=m_{\tilde{\tau}_L}$, whereas our framework favors $m_{\tilde{\tau}_R}<m_{\tilde{\tau}_L}$. A dedicated study of this distinction and its collider implications is left for future work.}

In Fig.~\ref{F5}, we display the neutralino--proton elastic scattering cross sections for both spin-independent (SI) and spin-dependent (SD) interactions. The upper panels show the SI scattering cross section, $\sigma_{\rm SI}$, for $\mu<0$ (left) and $\mu>0$ (right), while the lower panels present the corresponding SD scattering cross section, $\sigma_{\rm SD}$. For the $\mu<0$ case, a viable region persists in the $Z$-funnel that satisfies all imposed phenomenological constraints, including the relic density bound from {Planck2018} and the current direct-detection limits from {LZ} and {XENONnT}. The SI scattering rates in this region lie just below the projected {LZ} 1000-day sensitivity, indicating that this parameter space will be efficiently tested in the near future. In contrast, the Higgs-funnel solutions are already strongly constrained and are largely excluded by the current {LZ} bounds and its projected future reach. For $\mu>0$, both the $Z$- and Higgs-funnel regions are significantly more constrained, with the majority of solutions already excluded by present SI direct-detection limits. The suppression of $\sigma_{\rm SI}$ for $\mu<0$ can be understood from the structure of the neutralino--Higgs couplings. In the MSSM, the dominant SI contribution arises from $t$-channel exchange of the CP-even Higgs bosons $h$ and $H$. For negative $\mu$, these amplitudes may interfere destructively, particularly in the down-type quark sector, resulting in a substantial reduction of the effective neutralino-nucleon coupling \cite{Ellis:2000ds,Baer:2003jb}. By contrast, $\mu>0$ the interference is typically constructive, leading to an enhanced SI scattering rate and, consequently, stronger experimental constraints.

\begin{table}[h!]
	\centering
	\scalebox{0.87}{
		\begin{tabular}{|l|c|c|c|c|c|}
			\hline
			\hline
			&  Point 1 & Point 2  &  Point 3  & Point 4 & Point 5   \\
			\hline
			$m_{0}$        &  2648   & 2393   & 2357    & 2806 &  2951  \\
			$M_{2} $          & -2878   & -2587    & -2601  & -1087   & -1093    \\
			$M_{3}$         & 4892  & 4353   & 4271   & 1927  & 2032 \\
			$A_{0}/m_{0}$  &  -2.909 & -2.984 & -2.986   & -2.249  & -2.212   \\
			$\tan\beta$       & 32.18    & 29.62    & 30.18  &  39.06  & 39.33 \\
			$m_{H_d}$         &8989   & 9762   & 9742   &  8282  & 7362 \\
			$m_{H_u}$          & 7061   & 6682    & 6533  & 4481  & 4810 \\
			\hline 
            $\mu$                & {\bf -1848}   & {\bf -276}    & {\bf -413}   & {\bf 419}  & {\bf 352} \\
            \hline
			$m_h$                & {\bf 125.6}   & {\bf 125.5}    & {\bf 125.4}   & {\bf 125.5}  & {\bf 125.5} \\
			$m_H$                 & 8060 & 8954  & 8919  & 7075 & 6068  \\
			$m_A$                 & 8007 & 8895  & 8860  & 7028  & { 6028}    \\
			$m_{H^{\pm}}$         & 8060 & 8954  & 8919 & 7075 & { 6069} \\
			\hline
			$m_{\tilde{\chi}^0_{1,2}}$
			&  \textbf{80}, 1877 & \textbf{61},  288 & \textbf{ 44}, 427 & \textbf{ 44},  427& \textbf{ 61}, 360   \\
			$m_{\tilde{\chi}^0_{3,4}}$
			& 1883, 2557 & 291, 2291 & 430, 2300  & 436 ,972  & 368, 980 \\
			
			$m_{\tilde{\chi}^{\pm}_{1,2}}$
			&  1917, 2518 &  {297}, 2258& 441, 2268  & 405, 951  & 340, 957  \\
			\hline
			$m_{\tilde{g}}$ &  {9708}   &8724& 8565  & 4151 & 2528         \\
			$m_{ \tilde{u}_{L,R}}$
			& 1820, 8577 & 7964,7675 & 7831, 7530  & 4572, 4308  & 4782, 4610    \\
			$m_{\tilde{t}_{1,2}}$
			& 5832, 7262 & 5031, 6519 & 4938, 6400 & {1864}, 2957  & 2103, 3179   \\
			\hline $m_{ \tilde{d}_{L,R}}$
			&8820, 8677 &7964, 7849 & 7831, 7713  & 4573, 4585 & 4783, 4779    \\
			$m_{\tilde{b}_{1,2}}$
			& 7366, 7871 & 6535, 7157 & 6416, 7001  & 3009, 3354  & 3221, 3672 \\
			\hline
			$m_{\tilde{\nu}_{e,\mu}}$
			&6061   & 2649 & 2615 & 2661 & 2880     \\
			$m_{\tilde{\nu}_{\tau}}$   & 2253 & 1854 & 1769  & 1596 & 2031  \\
			\hline
			$m_{ \tilde{e}_{L,R}}$
			& 3077, 2910 & 2667, 2853 & 2633, 2836  & 2661, 3199  & 2890, 3197    \\
			$m_{\tilde{\tau}_{1,2}}$
			&  89, 2276 & 1057, 1880 & {875}, 1797  & 1137, 1626  & 1360, 2028   \\
			\hline
			
			$\sigma_{SI}({\rm pb})$
			&  $9.4\times 10^{-14}$& 2.3$\times10^{-11}$ & 2.6$\times 10^{-13}$ & 1.5$\times 10^{-11}$ & $4.1\times 10^{-11}$ \\

			$\sigma_{SD}({\rm pb})$
			&  1.0$\times 10^{-8}$ &2.1$\times 10^{-5}$ &4.1$\times 10^{-6}$   &4.0$\times 10^{-6}$  & $8.4 \times 10^{-6}$ \\
			$\Omega_{CDM}h^{2}$
			&  0.12189 & 0.12286 &0.12433  & 0.11453  & 0.11776 \\    $\mathcal{R}_{\tilde{\tau}_1}$& $0.112\%$   &  $16.19\%$&  $18.833\%$&  $--\%$  &$--\%$   \\
			\hline
			\hline
		\end{tabular}
	}
	\caption{Fundamental input parameters and the resulting sparticle mass spectra are presented for both signs of the $\mu$ parameter. Point 1 corresponds to the bulk region, Point 2 to the $H$-resonance, and Point 3 to the $Z$-resonance in the $\mu < 0$ scenario, while Point 4 represents the $Z$-resonance and Point 5 the $H$-resonance in the $\mu > 0$ scenario. All masses are given in units of GeV.}
	\label{table1}
\end{table}

The lower panels of Fig.~\ref{F5} show the corresponding SD cross sections. Here, the surviving $Z$-funnel solutions for $\mu<0$ remain below the current {LZ} sensitivity, but fall within the projected reach of future exposures. Unlike the SI case, the SD interaction is dominated by $Z$-boson exchange and is directly controlled by the Higgsino admixture of the lightest neutralino. Since the LSP in our viable parameter space is predominantly bino-like, with only a small Higgsino component, the $Z\tilde{\chi}_1^0\tilde{\chi}_1^0$ coupling is naturally suppressed, leading to correspondingly small values of $\sigma_{\rm SD}$. The SI and SD scattering cross sections, computed using \texttt{IsaTools}~\cite{Baer:2002fv}, are consistent with the current XENONnT~\cite{XENON:2023cxc} and LZ~\cite{LZ:2022lsv} constraints. However, a substantial part of the surviving Pati--Salam bulk and $Z$-funnel parameter space lies within the projected sensitivity of the 1000-day {LZ} run~\cite{LZ:2018qzl}, making these scenarios particularly testable in upcoming direct-detection searches. We note that the narrow-spectrum chargino--neutralino coannihilation region has already been extensively probed by existing direct-detection experiments.
The $Z$- and Higgs-funnel solutions correspond to the well-known resonant annihilation regime, where neutralino dark matter achieves the observed relic density through $s$-channel exchange of an on-shell or nearly on-shell mediator, satisfying $2m_{\tilde{\chi}_1^0}\simeq m_Z$ or $m_h$. In these regions, the lightest neutralino remains dominantly bino-like, with a moderate Higgsino admixture sufficient to generate the required annihilation rate while keeping both SI and SD scattering rates suppressed. Consequently, the invisible decay width constraints of the $Z$ and Higgs bosons are automatically satisfied throughout the viable parameter space. An important phenomenological implication of this analysis is that the surviving light-Higgsino solutions in the $Z$-pole region exhibit a clear preference for $\mu<0$, due to the destructive interference in the SI amplitude. By contrast, the Higgs-pole solutions do not show such a strong correlation with the sign of $\mu$, although the light-Higgsino regime is more readily realized for negative $\mu$.

Finally, in Table~\ref{table1} we present four representative benchmark points that summarize the main phenomenological features of our analysis. Point~1 corresponds to the bino–slepton bulk region, which satisfies both relic density requirements and current collider bounds while remaining within the yet-unexplored LHC parameter space. In this case, the LSP is a bino-dominated neutralino with a mass of approximately $0.080~\text{TeV}$, and the NLSP is identified as the lighter stau.
Point~2 corresponds to the Higgs-resonance annihilation channel, where the neutralino LSP is predominantly bino-like with a non-negligible higgsino admixture and has a mass of about $0.061~\text{TeV}$. Point~3 represents the $Z$-resonance regime in the $\mu < 0$ scenario, featuring a bino-like LSP with a small higgsino component and a mass near $0.044~\text{TeV}$. 
Similarly, Point~4 describes the $Z$-resonance annihilation channel in the $\mu > 0$ scenario, with a bino-dominated neutralino LSP of mass $\sim 0.040~\text{TeV}$, while Point~5 corresponds to the $H$-resonance case, again in the $\mu > 0$ scenario, with a bino-like LSP mass of approximately $0.061~\text{TeV}$. 
Taken together, these benchmark points span the phenomenologically viable regions of the supersymmetric Pati–Salam framework, including the bino-driven bulk regime and the Higgs- and $Z$-funnel annihilation channels.

\section{Conclusion}

We have studied the low-energy phenomenology of the Minimal Supersymmetric Standard Model (MSSM) derived from the supersymmetric
$SU(4)_C \times SU(2)_L \times SU(2)_R$ Pati--Salam framework, focusing on neutralino dark matter in the bulk annihilation and Higgs/$Z$-funnel regions. Using a comprehensive parameter-space scan subject to radiative electroweak symmetry breaking, a neutralino lightest supersymmetric particle (LSP), collider bounds, flavor constraints, and the \textit{Planck} relic-density bound, we identified the phenomenologically viable regions consistent with current experimental data. A central result of our analysis is the existence of a viable bulk annihilation region characterized by a bino-dominated neutralino LSP and a light right-handed stau as the next-to-lightest supersymmetric particle (NLSP). By imposing the conservative mass-splitting criterion
$\mathcal{R}_{\tilde{\tau}_1}\gtrsim10\%$,
we isolated genuine bulk solutions from coannihilation-dominated regimes and showed that the observed dark matter relic abundance can be achieved through conventional slepton-mediated annihilation. In this region, the spectrum is constrained to
$m_{\tilde{\chi}_1^0}\lesssim 110~{\rm GeV}$ and
$m_{\tilde{\tau}_1}\lesssim 120~{\rm GeV}$,
while scenarios involving a right-handed selectron NLSP are strongly disfavored by present LHC searches.
We also investigated the Higgs- and $Z$-funnel solutions for both signs of the Higgsino mass parameter $\mu$. Our results show that these resonance regions are strongly affected by current direct-detection constraints. In particular, for $\mu>0$, the Higgs- and $Z$-funnel solutions with light Higgsino admixture are severely constrained, primarily due to the enhancement of the spin-independent neutralino--nucleon scattering cross section. By contrast, for $\mu<0$, destructive interference in the Higgs-mediated scattering amplitude suppresses the spin-independent cross section, allowing a narrow but viable $Z$-funnel region to survive below the projected sensitivity of the 1000-day LZ exposure. An important implication of this study is that the negative-$\mu$ realization of the supersymmetric Pati--Salam framework remains particularly well-motivated in light of present collider and dark matter searches. The surviving parameter space predicts light electroweak states that are difficult to probe at the LHC because of compressed spectra and reduced production cross sections, but remain accessible to future lepton colliders such as CEPC and FCC-ee. At the same time, the predicted spin-independent and spin-dependent scattering rates place much of the viable parameter space within the reach of next-generation direct-detection experiments.
Taken together, our results demonstrate that the supersymmetric Pati--Salam framework continues to provide a predictive and experimentally testable realization of light neutralino dark matter, with the bulk and $Z$-funnel regions offering complementary discovery opportunities in future collider and dark matter searches.

\bibliographystyle{apsrev4-2}
\bibliography{refs}

@article{Witten:1981nf,
    author = "Witten, Edward",
    title = "{Dynamical Breaking of Supersymmetry}",
    reportNumber = "Print-81-0317 (PRINCETON)",
    doi = "10.1016/0550-3213(81)90006-7",
    journal = "Nucl. Phys. B",
    volume = "188",
    pages = "513",
    year = "1981"
}

@article{Dimopoulos:1981yj,
    author = "Dimopoulos, S. and Raby, S. and Wilczek, Frank",
    title = "{Supersymmetry and the Scale of Unification}",
    reportNumber = "NSF-ITP-81-31",
    doi = "10.1103/PhysRevD.24.1681",
    journal = "Phys. Rev. D",
    volume = "24",
    pages = "1681--1683",
    year = "1981"
}

@article{Amaldi:1991cn,
    author = "Amaldi, Ugo and de Boer, Wim and Furstenau, Hermann",
    title = "{Comparison of grand unified theories with electroweak and strong coupling constants measured at LEP}",
    reportNumber = "CERN-PPE-91-44, IEKP-KA-91-01",
    doi = "10.1016/0370-2693(91)91641-8",
    journal = "Phys. Lett. B",
    volume = "260",
    pages = "447--455",
    year = "1991"
}

@article{Ellis:1990wk,
    author = "Ellis, John R. and Kelley, S. and Nanopoulos, Dimitri V.",
    title = "{Probing the desert using gauge coupling unification}",
    reportNumber = "CERN-TH-5943-90, CTP-TAMU-97-90, ACT-19",
    doi = "10.1016/0370-2693(91)90980-5",
    journal = "Phys. Lett. B",
    volume = "260",
    pages = "131--137",
    year = "1991"
}

@article{Jungman:1995df,
    author = "Jungman, Gerard and Kamionkowski, Marc and Griest, Kim",
    title = "{Supersymmetric dark matter}",
    eprint = "hep-ph/9506380",
    archivePrefix = "arXiv",
    reportNumber = "SU-4240-605, UCSD-PTH-95-02, IASSNS-HEP-95-14, CU-TP-677",
    doi = "10.1016/0370-1573(95)00058-5",
    journal = "Phys. Rept.",
    volume = "267",
    pages = "195--373",
    year = "1996"
}

@inproceedings{Olive:2003iq,
    author = "Olive, Keith A.",
    title = "{TASI lectures on dark matter}",
    booktitle = "{Theoretical Advanced Study Institute in Elementary Particle Physics (TASI 2002): Particle Physics and Cosmology: The Quest for Physics Beyond the Standard Model(s)}",
    eprint = "astro-ph/0301505",
    archivePrefix = "arXiv",
    reportNumber = "UMN-TH-2127-03, TPI-MINN-03-02",
    pages = "797--851",
    month = "1",
    year = "2003"
}

@article{Drees:2005bx,
    author = "Drees, Manuel",
    editor = "Choi, Kiwoon and Kim, Jihn E. and Son, Dongchul",
    title = "{Neutralino Dark Matter in 2005}",
    eprint = "hep-ph/0509105",
    archivePrefix = "arXiv",
    doi = "10.1063/1.2149675",
    journal = "AIP Conf. Proc.",
    volume = "805",
    number = "1",
    pages = "48--54",
    year = "2005"
}

@article{Slavich:2020zjv,
    author = "Slavich, P. and others",
    editor = "Slavich, P. and Heinemeyer, S.",
    title = "{Higgs-mass predictions in the MSSM and beyond}",
    eprint = "2012.15629",
    archivePrefix = "arXiv",
    primaryClass = "hep-ph",
    reportNumber = "DESY 20-229, DESY-20-229, IFT-UAM/CSIC-20-184, FR-PHENO-2020-021, KA-TP-23-2020, MPP-2020-235, KA-TP-23-2020,
  MPP-2020-235, P3H-20-086, TTK-20-53, FERMILAB-PUB-21-575-T",
    doi = "10.1140/epjc/s10052-021-09198-2",
    journal = "Eur. Phys. J. C",
    volume = "81",
    number = "5",
    pages = "450",
    year = "2021"
}

@article{ATLAS:2012yve,
    author = "Aad, Georges and others",
    collaboration = "ATLAS",
    title = "{Observation of a new particle in the search for the Standard Model Higgs boson with the ATLAS detector at the LHC}",
    eprint = "1207.7214",
    archivePrefix = "arXiv",
    primaryClass = "hep-ex",
    reportNumber = "CERN-PH-EP-2012-218",
    doi = "10.1016/j.physletb.2012.08.020",
    journal = "Phys. Lett. B",
    volume = "716",
    pages = "1--29",
    year = "2012"
}

@article{CMS:2012qbp,
    author = "Chatrchyan, Serguei and others",
    collaboration = "CMS",
    title = "{Observation of a New Boson at a Mass of 125 GeV with the CMS Experiment at the LHC}",
    eprint = "1207.7235",
    archivePrefix = "arXiv",
    primaryClass = "hep-ex",
    reportNumber = "CMS-HIG-12-028, CERN-PH-EP-2012-220",
    doi = "10.1016/j.physletb.2012.08.021",
    journal = "Phys. Lett. B",
    volume = "716",
    pages = "30--61",
    year = "2012"
}

@article{You:2014vea,
    author = "You, Yi-Zhuang and Xu, Cenke",
    title = "{Interacting Topological Insulator and Emergent Grand Unified Theory}",
    eprint = "1412.4784",
    archivePrefix = "arXiv",
    primaryClass = "cond-mat.str-el",
    doi = "10.1103/PhysRevB.91.125147",
    journal = "Phys. Rev. B",
    volume = "91",
    number = "12",
    pages = "125147",
    year = "2015"
}

@article{Ananthanarayan:1992cd,
    author = "Ananthanarayan, B. and Lazarides, George and Shafi, Q.",
    title = "{Radiative electroweak breaking and sparticle spectroscopy with tan Beta approximately = m(t) / m(b)}",
    reportNumber = "BA-92-29, PRL-TH-92-16",
    doi = "10.1016/0370-2693(93)90361-K",
    journal = "Phys. Lett. B",
    volume = "300",
    pages = "245--250",
    year = "1993"
}

@inproceedings{Shafi:1991rs,
    author = "Shafi, Q. and Ananthanarayan, B.",
    title = "{Will LEP-2 narrowly miss the Weinberg-Salam Higgs boson?}",
    booktitle = "{Summer School in High-energy Physics and Cosmology}",
    reportNumber = "BA-91-76",
    pages = "233--244",
    month = "6",
    year = "1991"
}

@article{Hall:1993gn,
    author = "Hall, Lawrence J. and Rattazzi, Riccardo and Sarid, Uri",
    title = "{The Top quark mass in supersymmetric SO(10) unification}",
    eprint = "hep-ph/9306309",
    archivePrefix = "arXiv",
    reportNumber = "LBL-33997, UCB-PTH-93-15",
    doi = "10.1103/PhysRevD.50.7048",
    journal = "Phys. Rev. D",
    volume = "50",
    pages = "7048--7065",
    year = "1994"
}

@article{Ananthanarayan:1994qt,
    author = "Ananthanarayan, B. and Shafi, Q. and Wang, X. M.",
    title = "{Improved predictions for top quark, lightest supersymmetric particle, and Higgs scalar masses}",
    eprint = "hep-ph/9311225",
    archivePrefix = "arXiv",
    reportNumber = "BA-93-25-REV, PRL-TH-93-6-REV, BA-93-25, PRL-TH-93-6",
    doi = "10.1103/PhysRevD.50.5980",
    journal = "Phys. Rev. D",
    volume = "50",
    pages = "5980--5984",
    year = "1994"
}

@article{Rattazzi:1995gk,
    author = "Rattazzi, Riccardo and Sarid, Uri",
    title = "{The Unified minimal supersymmetric model with large Yukawa couplings}",
    eprint = "hep-ph/9505428",
    archivePrefix = "arXiv",
    reportNumber = "SU-ITP-94-16, RU-95-13",
    doi = "10.1103/PhysRevD.53.1553",
    journal = "Phys. Rev. D",
    volume = "53",
    pages = "1553--1585",
    year = "1996"
}

@article{Blazek:1996yv,
    author = "Blazek, Tomas and Carena, Marcela and Raby, Stuart and Wagner, Carlos E. M.",
    title = "{A Global chi**2 analysis of electroweak data in SO(10) SUSY GUTs}",
    eprint = "hep-ph/9611217",
    archivePrefix = "arXiv",
    reportNumber = "OHSTPY-HEP-T-96-026, CERN-TH-96-316, DESY-96-226",
    doi = "10.1103/PhysRevD.56.6919",
    journal = "Phys. Rev. D",
    volume = "56",
    pages = "6919--6938",
    year = "1997"
}

@article{Chkareuli:1998wi,
    author = "Chkareuli, J. L. and Gogoladze, I. G.",
    title = "{Unification picture in minimal supersymmetric SU(5) model with string remnants}",
    eprint = "hep-ph/9803335",
    archivePrefix = "arXiv",
    reportNumber = "CFP-IOP-98-01",
    doi = "10.1103/PhysRevD.58.055011",
    journal = "Phys. Rev. D",
    volume = "58",
    pages = "055011",
    year = "1998"
}

@article{Baer:2000jj,
    author = "Baer, Howard and Brhlik, Michal and Diaz, Marco A. and Ferrandis, Javier and Mercadante, Pedro and Quintana, Pamela and Tata, Xerxes",
    title = "{Yukawa unified supersymmetric SO(10) model: Cosmology, rare decays and collider searches}",
    eprint = "hep-ph/0005027",
    archivePrefix = "arXiv",
    reportNumber = "FSU-HEP-000216, UCCHEP-12-00, IFIC-00-18, FTUV-00504, UH-511-963-00, FTUV-000504",
    doi = "10.1103/PhysRevD.63.015007",
    journal = "Phys. Rev. D",
    volume = "63",
    pages = "015007",
    year = "2000"
}

@article{Baer:2010ny,
    author = "Baer, Howard and Kraml, Sabine and Lessa, Andre and Sekmen, Sezen and Tata, Xerxes",
    title = "{Effective Supersymmetry at the LHC}",
    eprint = "1007.3897",
    archivePrefix = "arXiv",
    primaryClass = "hep-ph",
    reportNumber = "LPSC10105, UH-511-1152-10",
    doi = "10.1007/JHEP10(2010)018",
    journal = "JHEP",
    volume = "10",
    number = "10",
    pages = "018",
    year = "2010"
}

@article{Ajaib:2013kka,
    author = "Ajaib, M. Adeel and Gogoladze, Ilia and Shafi, Qaisar",
    title = "{Sparticle Spectroscopy from SO(10) GUT with a Unified Higgs Sector}",
    eprint = "1307.4882",
    archivePrefix = "arXiv",
    primaryClass = "hep-ph",
    doi = "10.1103/PhysRevD.88.095019",
    journal = "Phys. Rev. D",
    volume = "88",
    number = "9",
    pages = "095019",
    year = "2013"
}

@inproceedings{Ajaib:2013uda,
    author = "Ajaib, M. Adeel and Gogoladze, Ilia and Shafi, Qaisar and Un, Cem Salih",
    title = "{Higgs and Sparticle Masses from Yukawa Unified SO(10): A Snowmass White Paper}",
    booktitle = "{Snowmass 2013}: {Snowmass on the Mississippi}",
    eprint = "1308.4652",
    archivePrefix = "arXiv",
    primaryClass = "hep-ph",
    month = "8",
    year = "2013"
}

@article{Gogoladze:2009ug,
    author = "Gogoladze, Ilia and Khalid, Rizwan and Shafi, Qaisar",
    title = "{Yukawa Unification and Neutralino Dark Matter in SU(4)(c) x SU(2)(L) x SU(2)(R)}",
    eprint = "0903.5204",
    archivePrefix = "arXiv",
    primaryClass = "hep-ph",
    doi = "10.1103/PhysRevD.79.115004",
    journal = "Phys. Rev. D",
    volume = "79",
    pages = "115004",
    year = "2009"
}

@article{Gogoladze:2009bn,
    author = "Gogoladze, Ilia and Khalid, Rizwan and Shafi, Qaisar",
    title = "{Coannihilation Scenarios and Particle Spectroscopy in SU(4)(c) x SU(2)(L) x SU(2)(R)}",
    eprint = "0908.0731",
    archivePrefix = "arXiv",
    primaryClass = "hep-ph",
    doi = "10.1103/PhysRevD.80.095016",
    journal = "Phys. Rev. D",
    volume = "80",
    pages = "095016",
    year = "2009"
}

@article{Gogoladze:2010fu,
    author = "Gogoladze, Ilia and Khalid, Rizwan and Raza, Shabbar and Shafi, Qaisar",
    title = "{$t-b-\tau$ Yukawa unification for $\mu < 0$ with a sub-TeV sparticle spectrum}",
    eprint = "1008.2765",
    archivePrefix = "arXiv",
    primaryClass = "hep-ph",
    doi = "10.1007/JHEP12(2010)055",
    journal = "JHEP",
    volume = "12",
    number = "12",
    pages = "055",
    year = "2010"
}

@article{Gomez:2020gav,
    author = "G{\'o}mez, Mario E. and Shafi, Qaisar and Un, Cem Salih",
    title = "{Testing Yukawa Unification at LHC Run-3 and HL-LHC}",
    eprint = "2002.07517",
    archivePrefix = "arXiv",
    primaryClass = "hep-ph",
    doi = "10.1007/JHEP07(2020)096",
    journal = "JHEP",
    volume = "07",
    number = "07",
    pages = "096",
    year = "2020"
}

@article{Djouadi:2022gws,
    author = "Djouadi, Abdelhak and Fonseca, Renato and Ouyang, Ruiwen and Raidal, Martti",
    title = "{Non-supersymmetric SO(10) models with Gauge and Yukawa coupling unification}",
    eprint = "2212.11315",
    archivePrefix = "arXiv",
    primaryClass = "hep-ph",
    doi = "10.1140/epjc/s10052-023-11696-4",
    journal = "Eur. Phys. J. C",
    volume = "83",
    number = "6",
    pages = "529",
    year = "2023"
}

@article{ATLAS:2019vcq,
    author = "{ATLAS Collaboration}",
    collaboration = "ATLAS",
    title = "{Search for squarks and gluinos in final states with jets and missing transverse momentum using 139 fb$^{-1}$ of $\sqrt{s}$ =13 TeV $pp$ collision data with the ATLAS detector}",
    reportNumber = "ATLAS-CONF-2019-040",
    journal = "ATLAS Conf. Note",
    month = "8",
    year = "2019"
}

@article{ATLAS:2020dsf,
    author = "Aad, Georges and others",
    collaboration = "ATLAS",
    title = "{Search for a scalar partner of the top quark in the all-hadronic $t{\bar{t}}$ plus missing transverse momentum final state at $\sqrt{s}=13$ TeV with the ATLAS detector}",
    eprint = "2004.14060",
    archivePrefix = "arXiv",
    primaryClass = "hep-ex",
    reportNumber = "CERN-EP-2020-044",
    doi = "10.1140/epjc/s10052-020-8102-8",
    journal = "Eur. Phys. J. C",
    volume = "80",
    number = "8",
    pages = "737",
    year = "2020"
}

@article{ATLAS:2019gdh,
    author = "Aad, Georges and others",
    collaboration = "ATLAS",
    title = "{Search for bottom-squark pair production with the ATLAS detector in final states containing Higgs bosons, $b$-jets and missing transverse momentum}",
    eprint = "1908.03122",
    archivePrefix = "arXiv",
    primaryClass = "hep-ex",
    reportNumber = "CERN-EP-2019-142",
    doi = "10.1007/JHEP12(2019)060",
    journal = "JHEP",
    volume = "12",
    number = "12",
    pages = "060",
    year = "2019"
}

@article{CMS:2019twi,
    author = "{CMS Collaboration}",
    collaboration = "CMS",
    title = "{Searches for new phenomena in events with jets and high values of the $M_{\mathrm{T2}}$ variable, including signatures with disappearing tracks, in proton-proton collisions at $\sqrt{s}=13~\mathrm{TeV}$}",
    reportNumber = "CMS-PAS-SUS-19-005",
    journal = "CMS PAS",
    year = "2019"
}

@article{CMS:2019xjf,
    author = "{CMS Collaboration}",
    collaboration = "CMS",
    title = "{Search for supersymmetry in proton-proton collisions at 13 TeV in final states with jets and missing transverse momentum}",
    reportNumber = "CMS-PAS-SUS-19-006",
    journal = "CMS PAS",
    year = "2019"
}

@article{Planck:2018nkj,
    author = "Aghanim, N. and others",
    collaboration = "Planck",
    title = "{Planck 2018 results. I. Overview and the cosmological legacy of Planck}",
    eprint = "1807.06205",
    archivePrefix = "arXiv",
    primaryClass = "astro-ph.CO",
    doi = "10.1051/0004-6361/201833880",
    journal = "Astron. Astrophys.",
    volume = "641",
    pages = "A1",
    year = "2020"
}

@article{Drees:1992am,
    author = "Drees, Manuel and Nojiri, Mihoko M.",
    title = "{The Neutralino relic density in minimal $N=1$ supergravity}",
    eprint = "hep-ph/9207234",
    archivePrefix = "arXiv",
    reportNumber = "SLAC-PUB-5860, DESY-92-101",
    doi = "10.1103/PhysRevD.47.376",
    journal = "Phys. Rev. D",
    volume = "47",
    pages = "376--408",
    year = "1993"
}

@article{King:2006tf,
    author = "King, S. F. and Roberts, J. P.",
    title = "{Natural implementation of neutralino dark matter}",
    eprint = "hep-ph/0603095",
    archivePrefix = "arXiv",
    doi = "10.1088/1126-6708/2006/09/036",
    journal = "JHEP",
    volume = "09",
    number = "09",
    pages = "036",
    year = "2006"
}

@article{XENON:2023cxc,
    author = "Aprile, E. and others",
    collaboration = "XENON",
    title = "{First Dark Matter Search with Nuclear Recoils from the XENONnT Experiment}",
    eprint = "2303.14729",
    archivePrefix = "arXiv",
    primaryClass = "hep-ex",
    doi = "10.1103/PhysRevLett.131.041003",
    journal = "Phys. Rev. Lett.",
    volume = "131",
    number = "4",
    pages = "041003",
    year = "2023"
}

@article{LZ:2022lsv,
    author = "Aalbers, J. and others",
    collaboration = "LZ",
    title = "{First Dark Matter Search Results from the LUX-ZEPLIN (LZ) Experiment}",
    eprint = "2207.03764",
    archivePrefix = "arXiv",
    primaryClass = "hep-ex",
    doi = "10.1103/PhysRevLett.131.041002",
    journal = "Phys. Rev. Lett.",
    volume = "131",
    number = "4",
    pages = "041002",
    year = "2023"
}

@article{LZ:2024zvo,
    author = "Aalbers, J. and others",
    collaboration = "LZ",
    title = "{Dark Matter Search Results from 4.2{\,}{\,}Tonne-Years of Exposure of the LUX-ZEPLIN (LZ) Experiment}",
    eprint = "2410.17036",
    archivePrefix = "arXiv",
    primaryClass = "hep-ex",
    reportNumber = "FERMILAB-PUB-24-0796-V",
    doi = "10.1103/4dyc-z8zf",
    journal = "Phys. Rev. Lett.",
    volume = "135",
    number = "1",
    pages = "011802",
    year = "2025"
}

@article{LZ:2018qzl,
    author = "Akerib, D. S. and others",
    collaboration = "LZ",
    title = "{Projected WIMP sensitivity of the LUX-ZEPLIN dark matter experiment}",
    eprint = "1802.06039",
    archivePrefix = "arXiv",
    primaryClass = "astro-ph.IM",
    reportNumber = "FERMILAB-PUB-18-054-AE-PPD",
    doi = "10.1103/PhysRevD.101.052002",
    journal = "Phys. Rev. D",
    volume = "101",
    number = "5",
    pages = "052002",
    year = "2020"
}

@article{Khan:2025ibo,
    author = "Khan, Imtiaz and Muhammad, Ali and Li, Tianjun and Raza, Shabbar and Pirzada and Khan, Mussawir",
    title = "{The Light Neutralino Dark Matter at Future Colliders in the MSSM with the Generalized Minimal Supergravity (GmSUGRA)}",
    eprint = "2509.23356",
    archivePrefix = "arXiv",
    primaryClass = "hep-ph",
    journal = "arXiv preprint",
    month = "9",
    year = "2025"
}

@article{Khan:2023ryc,
    author = "Khan, Imtiaz and Ahmed, Waqas and Li, Tianjun and Raza, Shabbar",
    title = "{Right-handed slepton bulk regions for dark matter in a generalized minimal supergravity}",
    eprint = "2312.07863",
    archivePrefix = "arXiv",
    primaryClass = "hep-ph",
    doi = "10.1103/PhysRevD.109.075051",
    journal = "Phys. Rev. D",
    volume = "109",
    number = "7",
    pages = "075051",
    year = "2024"
}

@article{FCC:2018byv,
    author = "Abada, A. and others",
    collaboration = "FCC",
    title = "{FCC Physics Opportunities}: {Future Circular Collider Conceptual Design Report Volume 1}",
    reportNumber = "CERN-ACC-2018-0056",
    doi = "10.1140/epjc/s10052-019-6904-3",
    journal = "Eur. Phys. J. C",
    volume = "79",
    number = "6",
    pages = "474",
    year = "2019"
}

@article{FCC:2018evy,
    author = "Abada, A. and others",
    collaboration = "FCC",
    title = "{FCC-ee: The Lepton Collider}: {Future Circular Collider Conceptual Design Report Volume 2}",
    reportNumber = "CERN-ACC-2018-0057",
    doi = "10.1140/epjst/e2019-900045-4",
    journal = "Eur. Phys. J. ST",
    volume = "228",
    number = "2",
    pages = "261--623",
    year = "2019"
}

@article{CEPCStudyGroup:2018ghi,
    author = "Dong, Mingyi and others",
    editor = "Guimar{\~a}es da Costa, Jo{\~a}o Barreiro and others",
    collaboration = "CEPC Study Group",
    title = "{CEPC Conceptual Design Report: Volume 2 - Physics {\&} Detector}",
    eprint = "1811.10545",
    archivePrefix = "arXiv",
    primaryClass = "hep-ex",
    reportNumber = "IHEP-CEPC-DR-2018-02, IHEP-EP-2018-01, IHEP-TH-2018-01",
    journal = "arXiv preprint",
    month = "11",
    year = "2018"
}

@article{Baer:1999sp,
    author = "Baer, Howard and Paige, Frank E. and Protopopescu, Serban D. and Tata, Xerxes",
    title = "{ISAJET 7.48: A Monte Carlo event generator for p p, anti-p, p, and e+ e- reactions}",
    eprint = "hep-ph/0001086",
    archivePrefix = "arXiv",
    reportNumber = "BNL-HET-99-43, FSU-HEP-991218, UH-511-952-00",
    journal = "arXiv preprint",
    month = "12",
    year = "1999"
}

@article{Hisano:1992jj,
    author = "Hisano, J. and Murayama, H. and Yanagida, T.",
    title = "{Nucleon decay in the minimal supersymmetric SU(5) grand unification}",
    eprint = "hep-ph/9207279",
    archivePrefix = "arXiv",
    reportNumber = "TU-400",
    doi = "10.1016/0550-3213(93)90636-4",
    journal = "Nucl. Phys. B",
    volume = "402",
    pages = "46--84",
    year = "1993"
}

@article{Yamada:1992kv,
    author = "Yamada, Youichi",
    title = "{SUSY and GUT threshold effects in SUSY SU(5) models}",
    reportNumber = "UT-626",
    doi = "10.1007/BF01650433",
    journal = "Z. Phys. C",
    volume = "60",
    pages = "83--94",
    year = "1993"
}

@article{Paige:2003mg,
    author = "Paige, Frank E. and Protopopescu, Serban D. and Baer, Howard and Tata, Xerxes",
    title = "{ISAJET 7.69: A Monte Carlo event generator for pp, anti-p p, and e+e- reactions}",
    eprint = "hep-ph/0312045",
    archivePrefix = "arXiv",
    journal = "arXiv preprint",
    month = "12",
    year = "2003"
}

@article{Baer:2016wkz,
    author = "Baer, Howard and Barger, Vernon and Gainer, James S. and Huang, Peisi and Savoy, Michael and Sengupta, Dibyashree and Tata, Xerxes",
    title = "{Gluino reach and mass extraction at the LHC in radiatively-driven natural SUSY}",
    eprint = "1612.00795",
    archivePrefix = "arXiv",
    primaryClass = "hep-ph",
    reportNumber = "UH-511-1269-16",
    doi = "10.1140/epjc/s10052-017-5067-3",
    journal = "Eur. Phys. J. C",
    volume = "77",
    number = "7",
    pages = "499",
    year = "2017"
}

@inproceedings{Allanach:2006fy,
    author = "Allanach, Benjamin C. and others",
    title = "{Les Houches physics at TeV colliders 2005 beyond the standard model working group: Summary report}",
    booktitle = "{4th Les Houches Workshop on Physics at TeV Colliders}",
    eprint = "hep-ph/0602198",
    archivePrefix = "arXiv",
    reportNumber = "FERMILAB-CONF-06-338-T, SLAC-PUB-11770",
    month = "2",
    year = "2006"
}

@article{ParticleDataGroup:2012pjm,
    author = "Beringer, J. and others",
    collaboration = "Particle Data Group",
    title = "{Review of Particle Physics (RPP)}",
    reportNumber = "SLAC-REPRINT-2014-001",
    doi = "10.1103/PhysRevD.86.010001",
    journal = "Phys. Rev. D",
    volume = "86",
    pages = "010001",
    year = "2012"
}

@article{ParticleDataGroup:2014cgo,
    author = "Olive, K. A. and others",
    collaboration = "Particle Data Group",
    title = "{Review of Particle Physics}",
    doi = "10.1088/1674-1137/38/9/090001",
    journal = "Chin. Phys. C",
    volume = "38",
    pages = "090001",
    year = "2014"
}

@article{LHCb:2012skj,
    author = "Aaij, R and others",
    collaboration = "LHCb",
    title = "{First Evidence for the Decay $B_s^0 \to \mu^+ \mu^-$}",
    eprint = "1211.2674",
    archivePrefix = "arXiv",
    primaryClass = "hep-ex",
    reportNumber = "CERN-PH-EP-2012-335, LHCB-PAPER-2012-043",
    doi = "10.1103/PhysRevLett.110.021801",
    journal = "Phys. Rev. Lett.",
    volume = "110",
    number = "2",
    pages = "021801",
    year = "2013"
}

@article{HFLAV:2012imy,
    author = "Amhis, Y. and others",
    collaboration = "HFLAV",
    title = "{Averages of B-Hadron, C-Hadron, and $\tau$-Lepton Properties as of early 2012}",
    eprint = "1207.1158",
    archivePrefix = "arXiv",
    primaryClass = "hep-ex",
    reportNumber = "SLAC-R-1002, FERMILAB-PUB-12-871-PPD",
    journal = "arXiv preprint",
    month = "7",
    year = "2012"
}

@article{HFLAV:2010pgm,
    author = "Asner, D. and others",
    collaboration = "HFLAV",
    title = "{Averages of $b$-hadron, $c$-hadron, and $\tau$-lepton properties}",
    eprint = "1010.1589",
    archivePrefix = "arXiv",
    primaryClass = "hep-ex",
    journal = "arXiv preprint",
    month = "10",
    year = "2010"
}

@article{Goldberg:1983nd,
    author = "Goldberg, H.",
    editor = "Srednicki, M. A.",
    title = "{Constraint on the Photino Mass from Cosmology}",
    reportNumber = "NUB-2592",
    doi = "10.1103/PhysRevLett.50.1419",
    journal = "Phys. Rev. Lett.",
    volume = "50",
    pages = "1419",
    year = "1983",
    note = "[Erratum: Phys.Rev.Lett. 103, 099905 (2009)]"
}

@article{Ellis:1983ew,
    author = "Ellis, John R. and Hagelin, J. S. and Nanopoulos, Dimitri V. and Olive, Keith A. and Srednicki, M.",
    editor = "Srednicki, M. A.",
    title = "{Supersymmetric Relics from the Big Bang}",
    reportNumber = "SLAC-PUB-3171",
    doi = "10.1016/0550-3213(84)90461-9",
    journal = "Nucl. Phys. B",
    volume = "238",
    pages = "453--476",
    year = "1984"
}

@article{Feng:2003zu,
    author = "Feng, Jonathan L.",
    editor = "Hewett, Joanne L. and Jaros, John and Kamae, Tsuneyoshi and Prescott, Charles",
    title = "{Supersymmetry and cosmology}",
    eprint = "hep-ph/0405215",
    archivePrefix = "arXiv",
    reportNumber = "UCI-TR-2004-7, SSI-2003-L11",
    journal = "eConf",
    volume = "C0307282",
    pages = "L11",
    year = "2003"
}

@article{Feng:2010gw,
    author = "Feng, Jonathan L.",
    title = "{Dark Matter Candidates from Particle Physics and Methods of Detection}",
    eprint = "1003.0904",
    archivePrefix = "arXiv",
    primaryClass = "astro-ph.CO",
    reportNumber = "UCI-TR-2009-13",
    doi = "10.1146/annurev-astro-082708-101659",
    journal = "Ann. Rev. Astron. Astrophys.",
    volume = "48",
    pages = "495--545",
    year = "2010"
}

@article{ATLAS:2019lng,
    author = "Aad, Georges and others",
    collaboration = "ATLAS",
    title = "{Searches for electroweak production of supersymmetric particles with compressed mass spectra in $\sqrt{s}=$ 13 TeV $pp$ collisions with the ATLAS detector}",
    eprint = "1911.12606",
    archivePrefix = "arXiv",
    primaryClass = "hep-ex",
    reportNumber = "CERN-EP-2019-242",
    doi = "10.1103/PhysRevD.101.052005",
    journal = "Phys. Rev. D",
    volume = "101",
    number = "5",
    pages = "052005",
    year = "2020"
}

@article{ATLAS:2020syg,
    author = "Aad, Georges and others",
    collaboration = "ATLAS",
    title = "{Search for squarks and gluinos in final states with jets and missing transverse momentum using 139 fb$^{-1}$ of $\sqrt{s}$ =13 TeV $pp$ collision data with the ATLAS detector}",
    eprint = "2010.14293",
    archivePrefix = "arXiv",
    primaryClass = "hep-ex",
    reportNumber = "CERN-EP-2020-166",
    doi = "10.1007/JHEP02(2021)143",
    journal = "JHEP",
    volume = "02",
    number = "02",
    pages = "143",
    year = "2021"
}

@article{ATLAS:2021kxv,
    author = "Aad, Georges and others",
    collaboration = "ATLAS",
    title = "{Search for new phenomena in events with an energetic jet and missing transverse momentum in $pp$ collisions at $\sqrt {s}$ =13  TeV with the ATLAS detector}",
    eprint = "2102.10874",
    archivePrefix = "arXiv",
    primaryClass = "hep-ex",
    reportNumber = "CERN-EP-2020-238",
    doi = "10.1103/PhysRevD.103.112006",
    journal = "Phys. Rev. D",
    volume = "103",
    number = "11",
    pages = "112006",
    year = "2021"
}

@article{ATLAS:2023xco,
    author = "{ATLAS Collaboration}",
    collaboration = "ATLAS",
    title = "{SUSY August 2023 Summary Plot Update}",
    reportNumber = "ATL-PHYS-PUB-2023-025",
    journal = "ATLAS Publ. Note",
    year = "2023"
}

@article{ATLAS:2019lff,
    author = "Aad, Georges and others",
    collaboration = "ATLAS",
    title = "{Search for electroweak production of charginos and sleptons decaying into final states with two leptons and missing transverse momentum in $\sqrt{s}=13$ TeV $pp$ collisions using the ATLAS detector}",
    eprint = "1908.08215",
    archivePrefix = "arXiv",
    primaryClass = "hep-ex",
    reportNumber = "CERN-EP-2019-106",
    doi = "10.1140/epjc/s10052-019-7594-6",
    journal = "Eur. Phys. J. C",
    volume = "80",
    number = "2",
    pages = "123",
    year = "2020"
}

@article{ATLAS:2022hbt,
    author = "Aad, Georges and others",
    collaboration = "ATLAS",
    title = "{Search for direct pair production of sleptons and charginos decaying to two leptons and neutralinos with mass splittings near the W-boson mass in $ \sqrt{s} $ = 13 TeV pp collisions with the ATLAS detector}",
    eprint = "2209.13935",
    archivePrefix = "arXiv",
    primaryClass = "hep-ex",
    reportNumber = "CERN-EP-2022-132",
    doi = "10.1007/JHEP06(2023)031",
    journal = "JHEP",
    volume = "06",
    number = "06",
    pages = "031",
    year = "2023"
}

@article{Lyu:2025wjn,
    author = "Lyu, Feng and Yuan, Jiarong and Cheng, Huajie and Wang, Jianxiong and Hameed, Rabia and Xu, Da and Zhuang, Xuai",
    title = "{Search potential for direct slepton pair production at the CEPC with ${ \sqrt{\boldsymbol s}}$ = 360 GeV}",
    eprint = "2501.03600",
    archivePrefix = "arXiv",
    primaryClass = "hep-ex",
    doi = "10.1088/1674-1137/ae2f4e",
    journal = "Chin. Phys. C",
    volume = "50",
    number = "3",
    pages = "033001",
    year = "2026"
}

@article{ATLAS:2018diz,
    author = "{ATLAS Collaboration}",
    collaboration = "ATLAS",
    title = "{Prospects for searches for staus, charginos and neutralinos at the high luminosity LHC with the ATLAS Detector}",
    reportNumber = "ATL-PHYS-PUB-2018-048",
    journal = "ATLAS Publ. Note",
    year = "2018"
}

@article{CMS:2018imu,
    author = "{CMS Collaboration}",
    collaboration = "CMS",
    title = "{Search for supersymmetry with direct stau production at the HL-LHC with the CMS Phase-2 detector}",
    reportNumber = "CMS-PAS-FTR-18-010, CMS-PAS-FTR-18-010",
    journal = "CMS PAS",
    year = "2018"
}

@article{Baer:2002fv,
    author = "Baer, Howard and Balazs, Csaba and Belyaev, Alexander",
    title = "{Neutralino relic density in minimal supergravity with coannihilations}",
    eprint = "hep-ph/0202076",
    archivePrefix = "arXiv",
    reportNumber = "FSU-HEP-020208",
    doi = "10.1088/1126-6708/2002/03/042",
    journal = "JHEP",
    volume = "03",
    number = "03",
    pages = "042",
    year = "2002"
}

@article{ATLAS:2014zve,
    author = "Aad, Georges and others",
    collaboration = "ATLAS",
    title = "{Search for direct production of charginos, neutralinos and sleptons in final states with two leptons and missing transverse momentum in $pp$ collisions at $\sqrt{s} =$ 8 TeV with the ATLAS detector}",
    eprint = "1403.5294",
    archivePrefix = "arXiv",
    primaryClass = "hep-ex",
    reportNumber = "CERN-PH-EP-2014-037",
    doi = "10.1007/JHEP05(2014)071",
    journal = "JHEP",
    volume = "05",
    number = "05",
    pages = "071",
    year = "2014"
}

@article{ATLAS:2023djh,
    author = "{ATLAS Collaboration}",
    title = "{Search for electroweak SUSY production in final states with two $\tau$-leptons in $\sqrt{s} = 13$ TeV $pp$ collisions with the ATLAS detector}",
    reportNumber = "ATLAS-CONF-2023-029",
    journal = "ATLAS Conf. Note",
    year = "2023"
}

@article{Ellis:2000ds,
    author = "Ellis, John R. and Ferstl, Andrew and Olive, Keith A.",
    title = "{Reevaluation of the elastic scattering of supersymmetric dark matter}",
    eprint = "hep-ph/0001005",
    archivePrefix = "arXiv",
    reportNumber = "CERN-TH-2000-001, UMN-TH-1834-2000, TPI-MINN-2000-01, CERN-TH-2K-001, TPI-MINN-00-01",
    doi = "10.1016/S0370-2693(00)00459-7",
    journal = "Phys. Lett. B",
    volume = "481",
    pages = "304--314",
    year = "2000"
}

@article{Baer:2003jb,
    author = "Baer, Howard and Balazs, Csaba and Belyaev, Alexander and O'Farrill, Jorge",
    title = "{Direct detection of dark matter in supersymmetric models}",
    eprint = "hep-ph/0305191",
    archivePrefix = "arXiv",
    reportNumber = "FSU-HEP-030515",
    doi = "10.1088/1475-7516/2003/09/007",
    journal = "JCAP",
    volume = "09",
    pages = "007",
    year = "2003"
}

@article{CidVidal:2018eel,
    author = "Cid Vidal, Xabier and others",
    editor = "Dainese, Andrea and Mangano, Michelangelo and Meyer, Andreas B. and Nisati, Aleandro and Salam, Gavin and Vesterinen, Mika Anton",
    title = "{Report from Working Group 3}: {Beyond the Standard Model physics at the HL-LHC and HE-LHC}",
    eprint = "1812.07831",
    archivePrefix = "arXiv",
    primaryClass = "hep-ph",
    reportNumber = "CERN-LPCC-2018-05",
    doi = "10.23731/CYRM-2019-007.585",
    journal = "CERN Yellow Rep. Monogr.",
    volume = "7",
    pages = "585--865",
    year = "2019"
}


\end{document}